\documentclass[useAMS,usenatbib]{mn2e}
\usepackage{graphicx}
\usepackage{amsmath}

\title[Dynamical behaviour of multiplanet systems close to 
their stability limit]
{Dynamical behaviour of multiplanet systems close to
their stability limit} 
\author[F. Marzari]{F. Marzari$^{1}$ \\
$^{1}$Dept. of Physics, University of Padova, 35131 Italy}
\begin{document}

\date{Accepted .....;  Received ..... ; in original form ........}

\pagerange{\pageref{firstpage}--\pageref{lastpage}} \pubyear{.....}

\maketitle

\label{firstpage}

\begin{abstract}
The dynamics of systems of two
and three planets, initially placed on circular 
and nearly coplanar orbits,  
is explored in the 
proximity of their stability limit.
The evolution of a large number of 
systems is numerically computed and their dynamical behaviour
is investigated  with the 
frequency 
map analysis as chaos indicator. Following the guidance of this 
analysis, it is found that for two--planet systems 
the dependence of the Hill limit on the planet mass,
usually made explicit through 
the Hill's radius parametrization, does not appear to be fully adequate. 
In addition, frequent cases of stable chaos are found in the proximity 
of the Hill limit. For three--planet systems, the usual approach
adopted in numerical explorations of their stability, where 
the planets are initially separated by multiples of the 
mutual Hill radius, 
appears too reducing. A detailed sampling of the 
parameter space reveals that systems with more packed inner planets
are stable well within previous estimates of the stability 
limit. This suggests that a two--dimensional approach is needed
to outline when three--planet systems are 
dynamically stable. 

\end{abstract}

\begin{keywords}
planetary systems; planets and satellites: dynamical evolution and stability
\end{keywords}

\section{Introduction}

Most observed extrasolar planets detected up to date move on
highly elliptical orbits, much larger than any solar 
system planet. The largest known eccentricity is 0.93, for 
HD 80606 b \citep{naef01}. 
This behaviour is at odds with 
formation theories predicting planets on circular orbits. 
Even migration by tidal interaction with the harboring disks
is characterized by dissipative interactions mostly leading to a reduction of
also initially small eccentricities. 
Different 
mechanisms may be invoked to excite
planet eccentricities \citep{namouni07}, but planet--planet scattering
appears to be the most valid candidate for exciting large orbital 
eccentricities \citep{rasio-ford96, weiden-marza96,
MW02,juritre08,naga08,
CHATTERJEE08, marzari10,
naga11,
benesv12}.
In a multi--planet system,  mutual gravitational perturbations 
between the planets may
cause an eccentricity growth leading to orbital crossing.
Eventually, the resulting dynamically violent chaotic phase 
is dominated by close encounters between the
planets ending when one or
more planets are ejected from the system on hyperbolic trajectories.
The initial orbital structure of the multi--planet system is
dramatically altered and a new stable configuration 
is reached with the surviving planets 
left on highly eccentric
non--interacting orbits possibly inclined respect to the initial
orbital plane. If tidal interaction
with the host star is invoked to circularize the orbit of the planet to the
periastron distance, this mechanism may explain a fraction 
of the observed 
``Hot Jupiters''.  In addition, it could also 
account for the production of pairs of planets in mean
motion resonances (\cite{ray08}).

A classical simple example of this scenario
is that of two jovian--size planets embedded in a 
circumstellar disk either trapped in resonance 
or evolving independently. Once 
the disk
is dispersed either by 
strong stellar
winds associated to the T--Tauri phase of the star
or by the progressive  
UV photo--evaporation induced by the central,
or nearby stars,
the gas damping 
on the planet orbits ceases and the system may become 
dynamically unstable undergoing a prolonged phase
of chaotic evolution characterized by repeated 
close encounters.  
Potentially, leftover planetesimals may 
still affect the planet evolution after the 
gas dissipation \citep{ray09} but here I focus on the 
phase where the only relevant forces are the mutual 
gravitational attractions between the planets.  
In this context, it is important 
to understand when a system of 2 or more planets,
even far from resonances, may become 
chaotic due to their mutual  perturbations. 
This may tell us how frequently planet--planet scattering
occurs on the basis of planet formation models. 

A significant effort has been devoted to 
outline a threshold value 
$\Delta$
for the initial separation between the planet orbits 
within which planet--planet scattering always occurs while 
beyond it the system is macroscopically stable and 
close encounters do not occur at least over an extended
timespan possibly comparable with the stellar age. 
The value of $\Delta$ has been extensively explored for 
systems of 2, 3 and more planets and lead to the introduction 
of the notion of "Hill stability". For a system of 2 planets, 
Hill--stable orbits are those for which the planets will conserve their 
ordering in terms of distance from the star. A more stringent 
formulation described in \cite{barnes-green06} and \cite{vermus13}
called 
"Lagrangian stability", requires that the outer planet cannot
escape from the system on a hyperbolic trajectory. As shown in
\cite{barnes-green06} these two last criteria give a
very  similar value
of the initial separation between the two planets.
In addition, the simulations performed by \cite{vermus13} show also that 
for initially circular orbits nearly all Hill stable systems are also
Lagrange stable. Since in this paper only orbits with initial 
small proper eccentricity are considered, 
we will adopt the Hill stability value $\Delta_h$ as 
estimate of the threshold value 
marking the transition from stable to unstable orbits.
However, while analytical estimated of $\Delta_h$ can be
found in the literature, the dynamics close to the Hill stability  limit 
has not been explored in details. Ad example, the orbits close 
to the Hill stability limit are stable in the sense of 
"quasi periodic" or are there cases of "stable chaos"?
Stable chaos is an oxymoron indicating orbits that exhibit
a chaotic evolution  but that, at the same time,
are  long--term macroscopically stable possibly because
the chaotic domain is thin. An example of this behaviour is provided
by asteroid (522) Helga as described in \cite{milani92}.

In this paper, by using the frequency map analysis
(hereinafter FMA), the 
stability properties of systems of two and three planets 
will be explored close to the Hill stability limit 
defined by an initial separation $\Delta_h$. 
If this limit marks a 
macroscopic transition between stability and instability,
one would expect a sudden drop of the frequency diffusion
speed.
This would mark the end of the resonance overlap region
so that beyond $\Delta_h$ strong instability 
will potentially develop only at individual low 
order resonances located farther out. 
We will see that this is not the case and 
that examples of "stable chaos" beyond $\Delta_h$ 
are frequently encountered. In 
addition, the FMA analysis will also show that the analytical 
estimate of 
$\Delta_h$ is not very precise. In particular, the dependence of 
$\Delta_h$ on the planet mass is not fully accounted 
by expressing it as a function of 
the Hill's radius.

In 
Sect. 2 the dynamics of two planets will be 
investigated close to the Hill's stability limit
by integrating a large number of systems 
and analysing their stability properties with the frequency 
map analaysis. In Sect. 3 the dynamics of 
systems with three planets will 
be analysed. It will be shown that the parametrization  
that uses multiples of the 
mutual Hill's radius to sample the initial parameter 
space and adopted in all            
previous stability studies misses many stable systems 
with an inner more packed planet pair. 

\section{Analytical and numerical estimates of $\Delta_h$}

For the case of two planets around a massive star, 
the elliptic three--body problem formulated in barycentric coordinates
leads to the following analytical estimate for $\Delta_h$ given
in implicit form \cite{glad93}:

\begin{equation}
\begin{split} 
& \alpha^{-3} \left(\mu_1 + \frac{\mu_2}{\left(1 + \frac{\Delta_h}{a_1}\right)^2}\right) (\mu_1 \gamma_1 + 
\mu_2 \gamma_2 \left(1 + \left(\frac{\Delta_h}{a_1}\right)^{1/2}\right)^2 \\
& > 1 + 3^{4/3} \frac {\mu_1 \mu_2} {\alpha^{4/3}} 
\end{split} 
\label{eq:gla1}
\end{equation}

where $\mu_i = m_i/M$, $\alpha = \mu_1 + \mu_2$, $\gamma_i = (1-e_i^2)^{1/2}$,
$m_i$ is the mass of planet $i$, $M$ is the mass of the star, $e_i$ is the
eccentricity of planet $i$.
In the case of initially circular (and coplanar) orbits
the above equation can be solved and,
to the lowest order in mass, it gives:
\begin{equation}
\Delta_h \sim 2.40 (\mu_1 +\mu_2)^{1/3} a_1
\label{eq:gla2}
\end{equation}

When the eccentricities and inclinations of the planets are not small,
more complex expressions can be derived \citep{donnison06} but we 
will concentrate in this paper on the more simple case 
of nearly coplanar and initially circular orbits. The above formula has been 
numerically tested by \cite{CH96} showing that a more precise 
value is given by:

\begin{equation}
\Delta_h \sim 2 \sqrt{3} R_H
\label{eq:cha1}
\end{equation}

where $R_H$ is the planets' mutual Hill radius given by

\begin{equation}
R_H = \left(\frac{m_1 + m_2} {3 M_{\odot}}\right )^{1/3}
\left( \frac{\left(a_1 + a_2\right)} {2}\right)
\label{eq:cha2}
\end{equation}

For systems of three planets, numerical integrations of orbits 
have been performed showing that the timescale for the 
onset of instability grows logarithmically by increasing 
the mutual separation between the planets \citep{MW02, CH96,
CHATTERJEE08}. This has been interpreted as a decrease of the 
chaotic diffusion which should finally lead to stable quasi 
periodic orbits beyond a threshold value of 
separation. These studies are based on a parametrization 
of the initial separation between the planets based on multiples of 
the Hill radius. Once fixed the inner planet
semimajor axis, the next one is
placed at:

\begin{equation}
a_{i+1}  = a_i + K R_H
\label{eq:cha3}
\end{equation}

In all these numerical simulations it was found that,
whenever a third planet is added, the Hill-stable region is 
placed beyond the $\Delta_h$ value computed 
independently for each 
pair of planets. In addition, the value of $K$ defining 
the $\Delta_h$ for which systems are stable 
depends on the mass of the planets. 

\section[]{Use of the FMA to detect chaotic behaviour}

There is a variety of numerical tools useful 
for the detection of chaos. The computation of the 
Maximum Lyapunov Exponent (MLE) which measures the 
exponential divergence of close orbits \citep{benettin76} 
has the drawback of being difficult to be implemented in an automatic way  
for exploring a large sample of orbits. More suited for this task 
is the computation of the Mean Exponential Growth Factor 
of Nearby Orbits (MEGNO) introduced by 
\citep{cincotta-simo00} and used ad example by 
\cite{god02,god03}, or the Frequency Map Analysis
that is adopted in this paper. 
The FMA (Frequency Map Analysis) \citep{lask93},
implemented as described in \cite{marz03}, is used to measure the
diffusion in the phase space of the fundamental frequencies of 
a dynamical system. 
The algorithm of the FMA method computes
the strongest peak in the Fourier transform of a
time series with the MFT (Modified Fourier Transform)
high precision method \citep{lask93,
sine96}. For a system of 2 planets we consider the non--singular variables
$h,k$ in their complex--number
form, and with the MFT we compute one of the 
two main frequencies 
$g_i$ predicted 
by the traditional Laplace--Lagrange theory 
\citep{murray-dermottSS}.   They are the eigenvalues 
of the 2x2 matrix $A$ function only of the masses and 
semimajor axes of the two planets whose elements are given as:

\begin{align}
        A_{jj}&  =  \frac {n_j} {4} \frac {m_k} {M_{\odot}+m_j}
                    \alpha_{12}^2 b^{(1)}_{3/2} 
                    (\alpha_{12}) \\
        A_{jk}&  =  -\frac {n_j} {4} \frac {m_k} {M_{\odot}+m_j}
                    \alpha_{12}^2 b^{(2}_{3/2} 
                    (\alpha_{12})  , \label{secular}
\end{align}

where $j=1,2$, $k=2,1$, $j \ne k$ and 
$\alpha_{12} = a_1/a_2$ since in our case $a_1 \le a_2$.  
The $b^{(1)}_{3/2}(\alpha)$  and 
$b^{(2}_{3/2}(\alpha)$ are the Laplance coefficients whose 
expression can be found in \cite{murray-dermottSS}.
The precise value  
of one of the two frequencies $g_i$ 
(the stronger one in the power spectrum of
the inner planet) is computed over running
windows covering the whole timespan of the 
numerical integration of the planet orbits.
Its standard deviation $\sigma_g$ 
is hence used to compute a 
chaos indicator $c_v = log(\sigma_g)/g$.  The system is
quasi--periodic only if the proper frequency is not
changing with time, otherwise it is chaotic \citep{lask93}
and $c_v$ measures the variation in time of the proper
frequency and then the diffusion speed of the orbit
in the phase space.

\subsection[]{The FMA set up}

To investigate the dynamics of a two--planet system
close to the Hill stability limit,  
the initial semimajor axis of the inner one  $a_1$ is fixed at 3 AU 
while $a_2$ ranges  from 3 AU to 
6 AU. The initial eccentricity is set to 0 for both planets
while the mutual inclination is randomly selected in 
the range $[0^o,5^o]$, a reasonable initial value for 
planets formed within a protostellar disk. 
The other orbital angles are all randomly 
selected between $[0^o,360^o]$. The masses of the planets 
are all equal and varied from $5.1 \times 10^{-5} M_{\odot}$ 
(Neptune--size planets) to  $2.8 \times 10^{-4} M_{\odot}$
(Saturn--size planets) and to
$9.5 \times 10^{-4} M_{\odot}$ (Jupiter--size planets).
The trajectories of the bodies are integrated over 
10 Myr and the FMA is applied to the $h,k$ variables 
of the inner planet over running windows 
of 2 Myr. The symplectic integrator SyMBA \citep{levdun94,levdun98} 
is used with a timestep of 10 days and the output is recorded
every 5 yrs.  In this version of SyMBA, whenever two bodies 
suffer a mutual encounter the time step for
the involved bodies is recursively subdivided
to whatever level is required. However, it must be noted that 
the numerical integration is halted after the first close 
encounter between planets since it is a clear indication 
of instability and the FMA analysis is not needed in this 
case. The timestep 
of 10 days adopted in the simulations 
is approximately 1/190 of the orbital period of 
the inner planet and, as a consequence, the numerical algorithm 
can comfortably handle the eccentricity built up 
by the dynamical instability that leads to the first
close approach. 

\section[]{Stability of two--planet systems}

\begin{figure}
\hskip -0.3 truecm
\resizebox{90mm}{!}{\includegraphics[angle=-90]{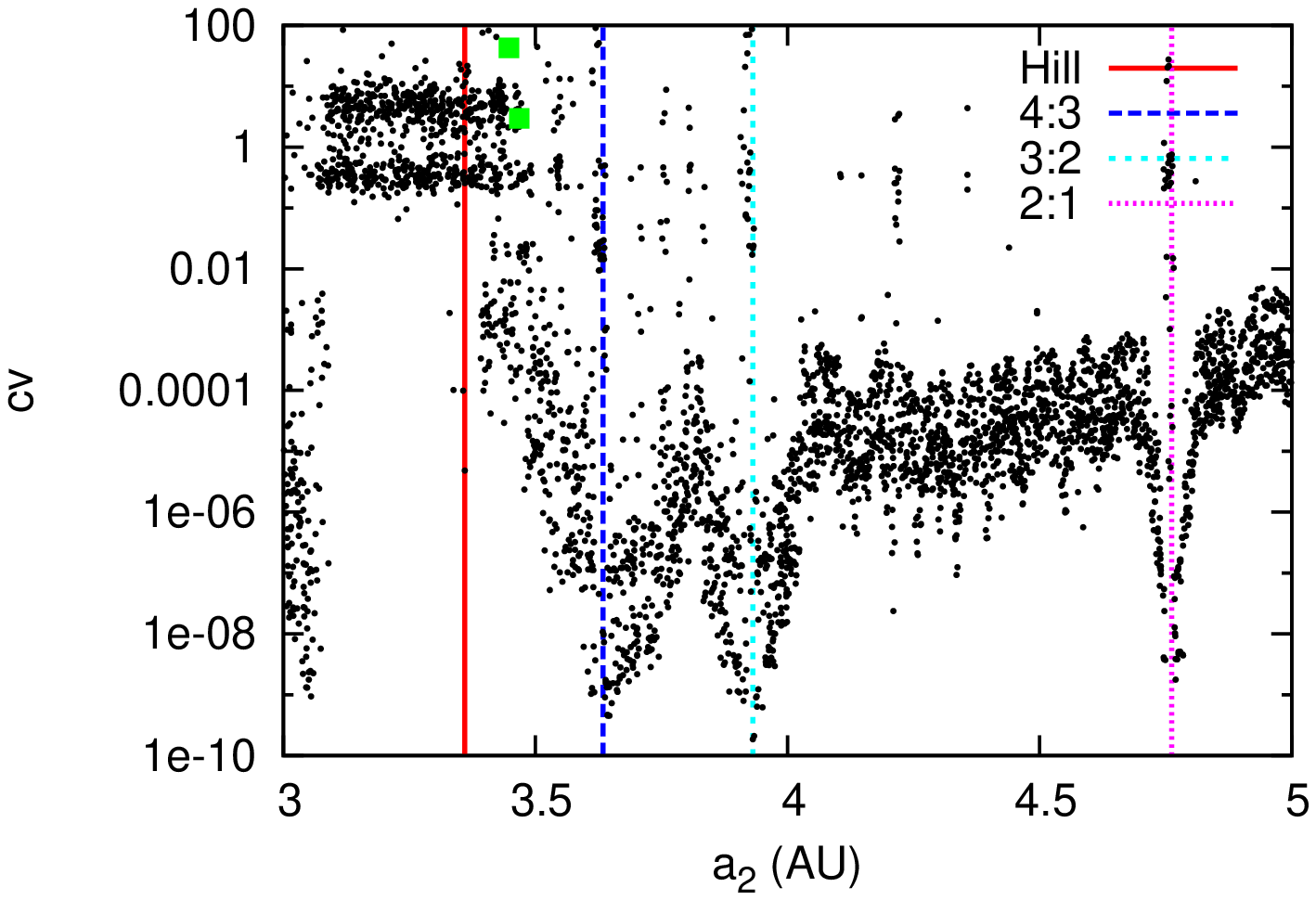}}
\resizebox{90mm}{!}{\includegraphics[angle=-90]{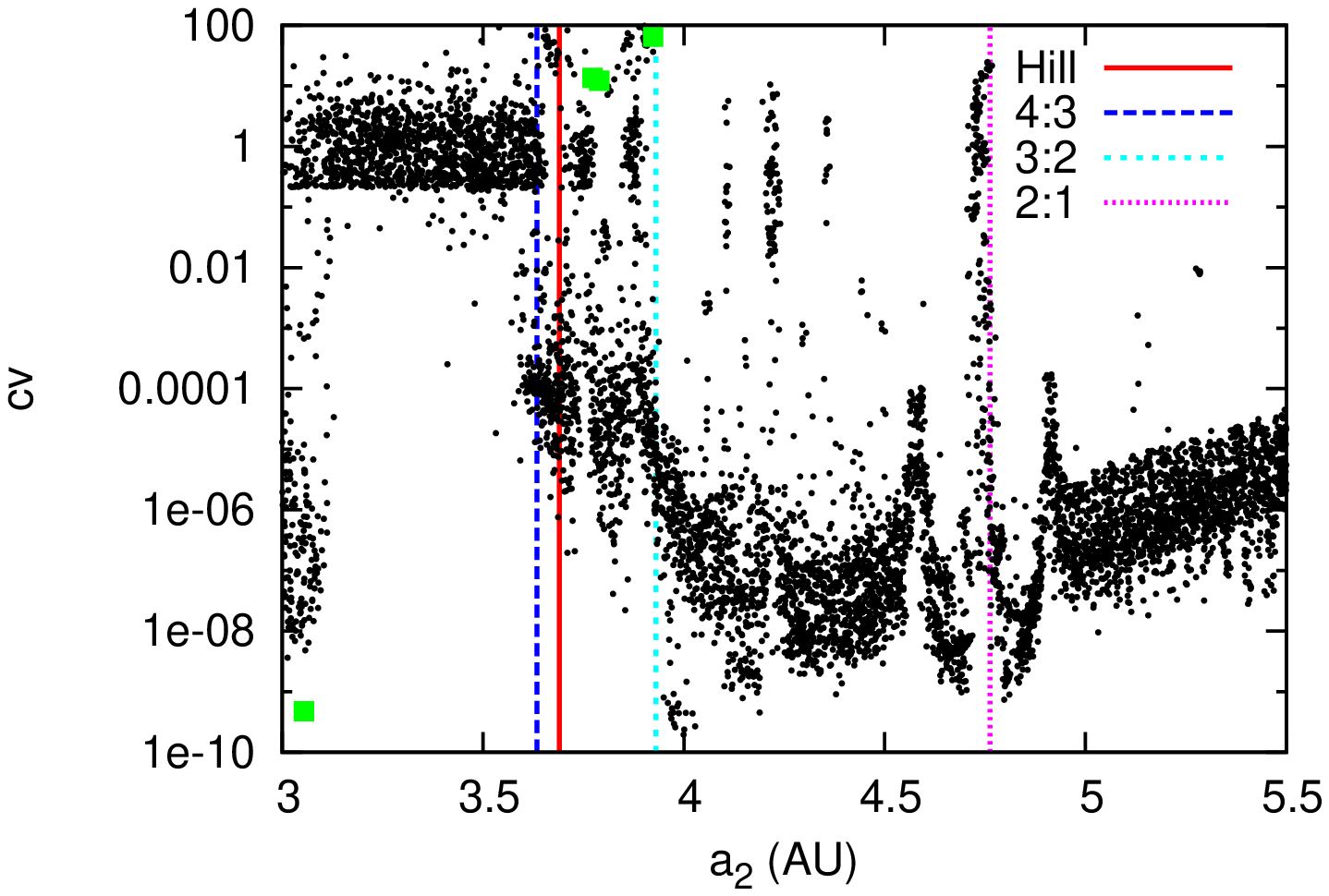}}
\resizebox{90mm}{!}{\includegraphics[angle=-90]{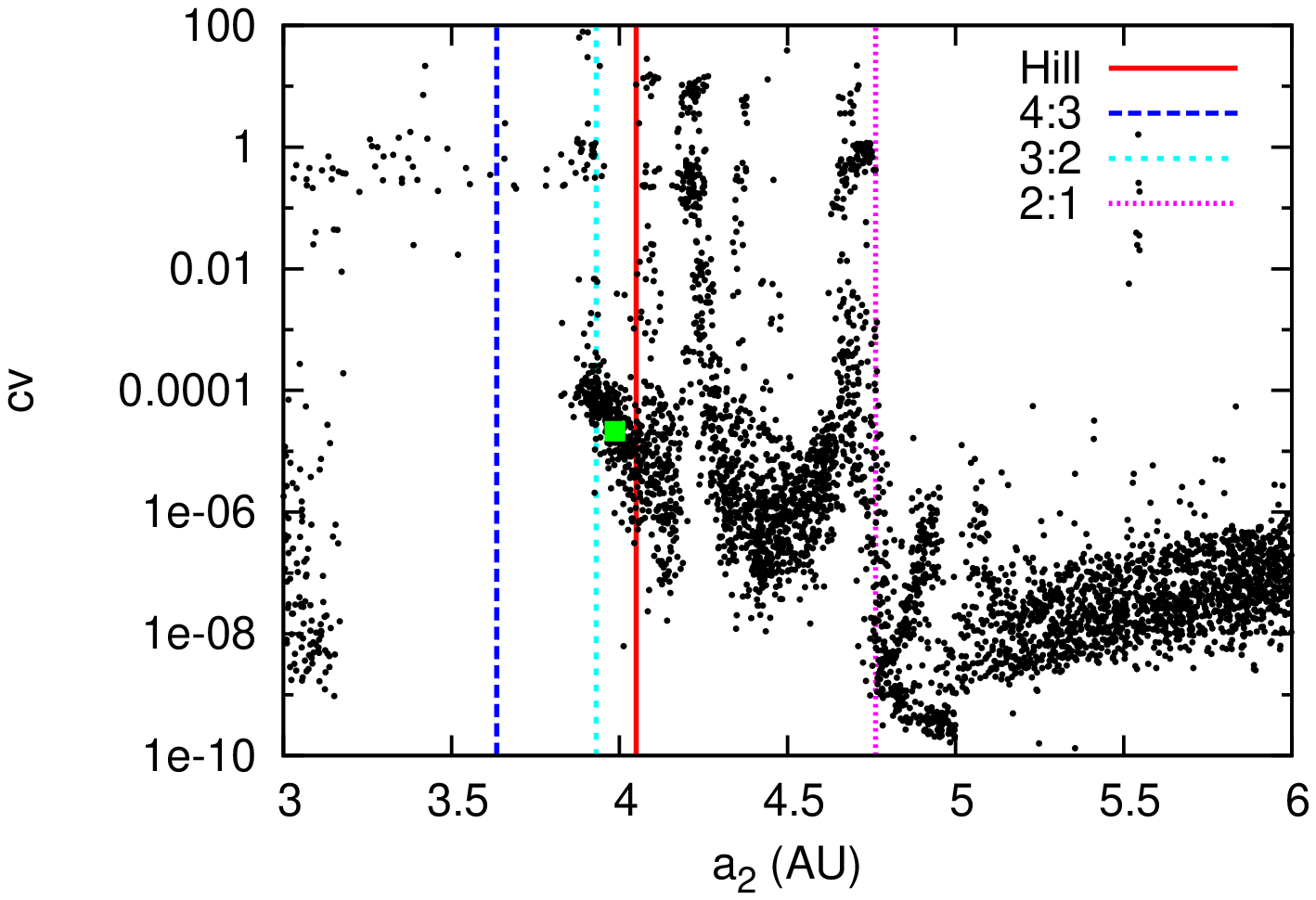}}
\caption{Values of the FMA chaos indicator $c_v$ for different 
values of the second planet semimajor axis $a_2$. The 
vertical lines mark the location of the Hill stability 
limit, estimated by $\Delta_h$, and of the main first
order mean motion resonances. 
The green squared points indicate those cases that have been 
integrated over 5 Gyr to explore their long
term dynamical behaviour. 
In the top panel the mass of the two planet is 
$5.1 \times 10^{-5} M_{\odot}$
(Neptune--size planets), in the middle panel  
the planet mass is set to $2.8 \times 10^{-4} M_{\odot}$
(Saturn--size planets) and in the bottom panel the mass is 
$9.5 \times 10^{-4} M_{\odot}$ (Jupiter--size planets).
}
\label{f1}
\end{figure}

In Fig.\ref{f1} the three panels show the outcome of 
the FMA analysis for the different values of the planet masses. 
The inner planet is on a fixed orbit at 3 AU so we plot
the value of $c_v$, the chaos indicator, 
as a function of the outer planet 
semimajor axis $a_2$. 
One would expect a sudden drop of the stability index $c_v$ 
beyond the Hill stability limit but this occurs only partly
and at a slightly different value respect to the analytical 
prediction $\Delta_h$ (Eq.~(\ref{eq:cha2}). The mismatch depends 
on the  
mass of the planets. 

For Neptune--size planets,  $\Delta_h$ 
is underestimated and the decrease in the stability 
properties shown by the FMA analysis  
begins farther out. This is confirmed by the long
term integration of two cases,
marked by green squares in Fig.\ref{f1}, top panel,
which have 
an initial separation larger than $\Delta_h$. 
Both systems become 
macroscopically chaotic before 1 Gyr and 
undergo close encounters. This indicates that the 
Hill's stability limit is farther out respect to the analytical 
prediction. The minimum value of $c_v$ is reached 
at about 3.7 AU, just after the 4:3 mean motion resonance 
(hereinafter MMR). Beyond the 4:3, unstable orbits are found 
only at 
individual and well separated 
mean motion resonances like, ad example, the 2:1.  
From the analysis of $c_v$, a good choice for an initial 
estimate of $\Delta$, marking the transition 
between stable and unstable orbits, could be the 
separation between the location of the inner planet and the 
4:3 resonance. In this case $\Delta$ would be about 0.63 AU 
against the value of $\Delta_h$ that is 0.36 AU. 

\begin{figure}
\hskip -0.8 truecm
\resizebox{90mm}{!}{\includegraphics[angle=0]{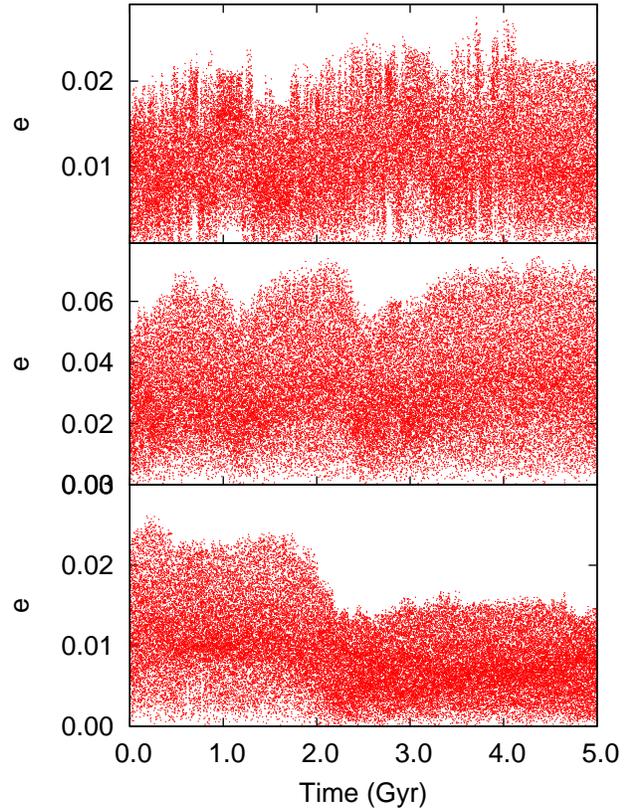}}
\caption{Long term evolution of 3 cases 
($a_2 $ = 3.77, 3.79, and 3.92) of stable chaos 
highlighted in Fig.\ref{f1} (middle panel) as green squares. The 
eccentricity evolution of the inner planet confirms 
the chaotic character of the systems in spite of their
macroscopic stability over 5 Gyr.  
}
\label{f2}
\end{figure}

For Saturn mass planets (Fig.\ref{f1} middle panel), 
the value of $\Delta_h$ appears
to better outline the transition from stability to instability. 
It is very close to the 4:3 MMR and the drop in the value of 
$c_v$ begins very close to it. However, beyond $\Delta_h$
a consistent fraction of planet pairs have still a large 
value of $c_v$ until the 3:2 MMR is
encountered. Once the resonance is passed, a minimum of 
$c_v$ is met and it remains small up to the 2:1 resonance
and beyond, apart from unstable orbits at lower order 
mean motion resonances. The region between the Hill stability
boundary and the 3:2 MMR is populated by cases of 
"stable chaos". The irregular motion is limited in 
scale and it does not lead to large eccentricities,
close encounters and then macroscopic chaotic evolution. 
Their behaviour is clearly illustrated in Fig.\ref{f2} where
the long term evolution of the eccentricity of 
the inner planet of a few selected systems located
between the Hill stability
boundary and the 3:2 MMR is shown. For these systems 
it looks like the chaotic domain, 
which is due to overlap of multiple resonances, 
does not contain a region 
where the planets may increase their eccentricities 
by mutual perturbations up to the point of triggering
close encounters. 
 
In the case of Jupiter--size planets (Fig.\ref{f1} bottom panel),
the dynamics appears to be complex. Only a few cases 
are stable within the 3:2 MMR beyond which the 
$c_v$ quickly drops. 
The value of 
$\Delta_h$ overestimates the transition from stable 
to unstable systems and at smaller separations
there are cases that are not chaotic and 
stable over 5 Gyr. 
However, this is not an effect due to the presence of 
the 3:2 MMR. \cite{bargre07}  have shown that within a MMR the stability
properties of planetary systems may be increased. A similar 
result was obtained by \cite{mst06}  for the 2:1 resonance where 
long term stable and chaotic zones can be found.  
In the case of Jupiter-pair systems,  the decrease of the stability 
properties begin outside the 3:2 MMR and the system marked
by the green square in the bottom panel of Fig.\ref{f1} is 
not in a 3:2 MMR (no libration or slow circulation 
of the critical arguments is observed) and it is stable over 5 Gyr. 
As a consequence, the analytical estimate of 
$\Delta_h$ indeed overestimates the location of the transition
from stable to unstable orbits. In addition, the 
stability marker $c_v$ in the bottom panel of Fig.\ref{f1} 
decreases away from the location of the 3:2 MMR while, usually, 
getting closer to the separatrix leads to systems with lower 
values of $c_v$.

The green square in Fig.\ref{f1},
bottom panel, marks a case which is stable over 5 Gyr 
in spite of being located within $\Delta_h$. The evolution 
of the eccentricity of both planets is periodic and it does 
not even show signs of stable chaos. For Jupiter--size planets,  
the location of the 3:2 MMR is a better indicator
of stability respect to $\Delta_h$. The region between 
the 3:2 and 2:1 MMR are characterized by the presence of 
both unstable and "stable chaos" orbits in particularly 
close to the 7:4 MMR. 

In conclusion, the transition between stable and unstable orbits
in two--planet systems appears to be more complicated than 
thought. The accuracy of the 
semi--empirical estimate of the initial separation 
$\Delta_h$ depends on the mass of 
the planets in the system. In addition, the transition does not
occur as a sudden change from 
chaotic to non--chaotic orbits since cases of 
stable chaos are found close to the threshold value.   
It must also be noted that the offset of $\Delta_h$ 
respect to the real stability limit observed in 
Fig.\ref{f1}
cannot be ascribed to the 
difference between the definition of Hill 
stability and Lagrange stability. The main reason is that 
the mismatch between the semi--analytical estimate of 
$\Delta_h$ and the real beginning of instability has a systematic
trend ranging from $\Delta_h$ being too small for Neptune--size
planets to being too large for Jupiter--size planets while 
it is a very good approximation for Saturn--size planets. 
This trend 
does not reflect the fact that the Lagrange
stability limit is more restrictive being then either coincident
or systematically larger than $\Delta_h$. Moreover, as stated 
by \cite{barnes-green06} and \cite{vermus13}, the two limits 
for systems with initially circular orbits practically 
coincide.

It is noteworthy that, for any value of the planet masses
considered here, there 
are cases of stable Trojan planet configurations. Ad example, in 
the case of Jupiter--size planet pairs
long term numerical simulations of a few sample cases 
have shown 
that a low value of
$c_v$ implies dynamical stability over 5 Gyr. 
The possibility that extrasolar planets may be 
trapped in a 1:1 resonance has been 
explored in details in previous publications \citep{dvorak07,had09}.
In particular,  \cite{laughchamb02} have shown that 
Trojan--type orbits are stable for a mass ratio
$\mu = \frac {(m_1 + m_2)} {(m_1+m_2+M_{\odot})} \leq 0.03812$
while the condition for horseshoe orbit stability is $\mu \leq 
0.0004$. Hence, we expect that only among Neptune and Saturn--size planets
(top and middle panels of Fig.\ref{f1})  
stable pairs of planets 
in horseshoe orbits
can be found.

In all 3 panels of Fig.\ref{f1}, after a marked decrease 
of $c_v$ in the proximity of $\Delta_h$, 
its value appears to slowly increase again for large values of 
$a_2$. This growing trend 
is due to a degradation of the precision 
in measuring the variation of the frequencies $g_1$ and 
$g_2$. This is related to the decrease of both 
$g_1$ and $g_2$ for a larger separation between the 
planets. Since the time span of the integration and of 
the running windows where the frequencies are computed is constant, 
less periods of the $h,k$ variables are covered 
for larger values of $a_2$ and, as a consequence,  the 
precision decreases. At the same time, the forced eccentricity
of each planet is smaller for larger values of $a_2$ and 
the $h,k$ variables become progressively smaller, 
increasing the numerical error in the computation of the
associated frequencies. This does not affect the overall outcome
of the FMA analysis, since this study is focused on what  
happens close to the value of $\Delta_h$ and in its 
surroundings.

\section[]{The case of three equal--mass planets}

Observed systems Ups And, 55 Cnc, and HD 37124 have proven 
that it is indeed possible to form more than two giant planets around 
solar type stars.
A system with three giant planets has a more rich and 
complex dynamical behaviour compared to two planet systems. 
The onset of instability depends on a wider parameter
space and its outcome is dynamically more various. 
It has been a custom in previous studies of dynamical scattering of 
systems of 3 or more planets 
\citep{CH96, MW02, CHATTERJEE08}
to explore the stability limit
using the parametrization described by 
Eq.~(\ref{eq:cha3})  with $K$ depending on
the planet masses (assuming they are all equal).  
The initial semimajor axis of each of the three planets are usually 
selected according to the relations $a_2 = a_1 + K R_{H1,2}$
and $a_3 = a_2 + K R_{H2,3}$.
However, the problem has an intrinsic higher dimensionality and 
this parametrization, as it will be shown here, 
appears inadequate to fully describe 
the stability properties of three planets and to 
derive where the 
transition between stable and unstable systems occurs. 

As for two planets, to explore the dynamics of three--planet 
systems close to their
stability threshold, we integrated the orbits 
of a large number (about $1 \times 10^5$) 
of putative systems and computed, 
via the FMA, the $c_v$ coefficient. 
In this case, 
we have
randomly sampled both 
$a_2$ and $a_3$, the initial semimajor axis of the second 
and third 
planet. The initial eccentricity is set to 0 for all
three planets  and their
inclination is sampled in the same range of the 2--planet cases. 
All other angles are randomly chosen. 
The FMA is performed on the $h,k$ variables of the inner planet. 
To illustrate the behaviour of the 
chaos indicator $c_v$ it is necessary in this case to resort 
to color mapping
plots where the semimajor axis of the two outer planets are 
adopted as variables. 

\begin{figure}
    \centering
    \includegraphics[width=0.40\textwidth,angle=-90]{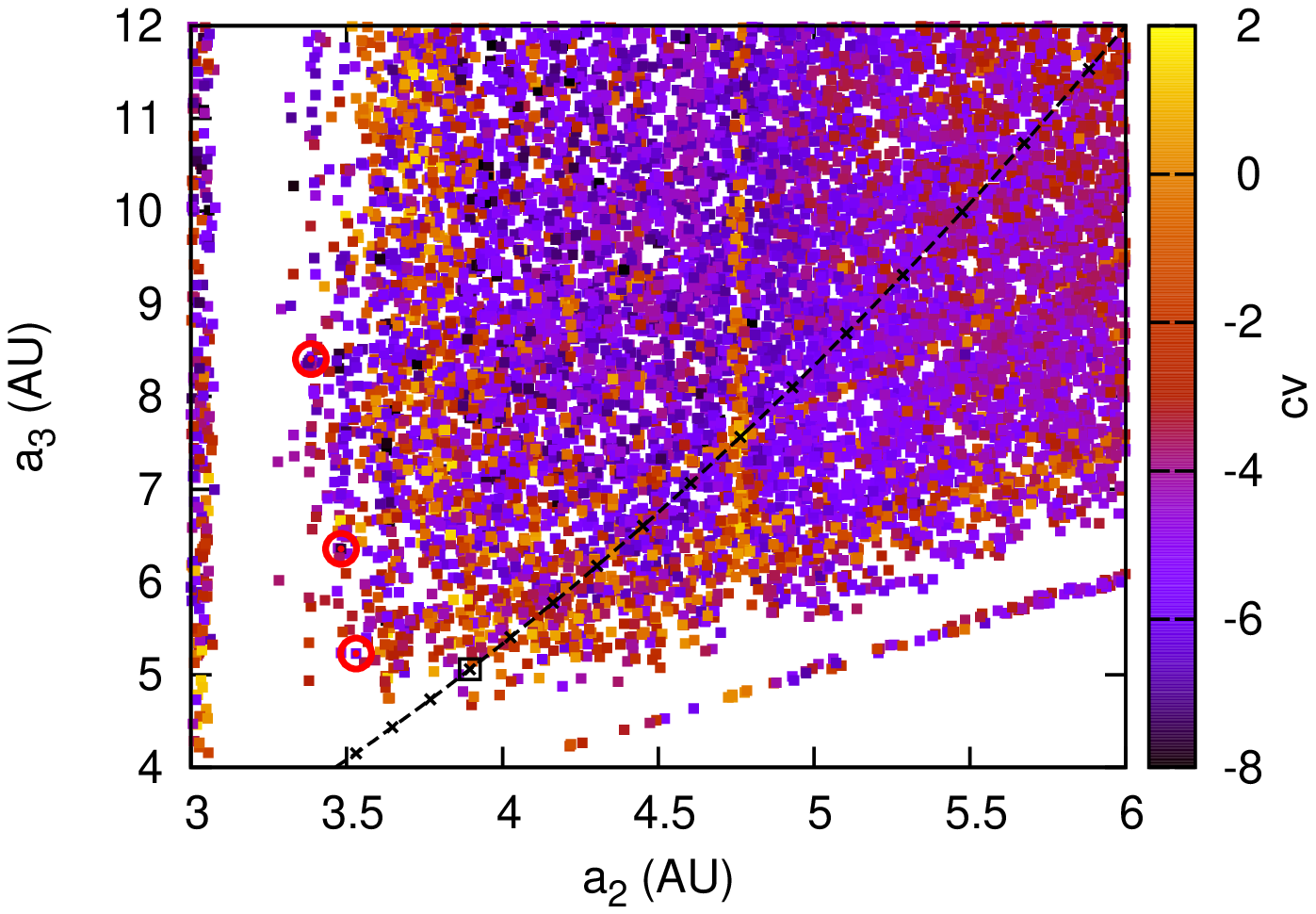}
    \includegraphics[width=0.40\textwidth,angle=-90]{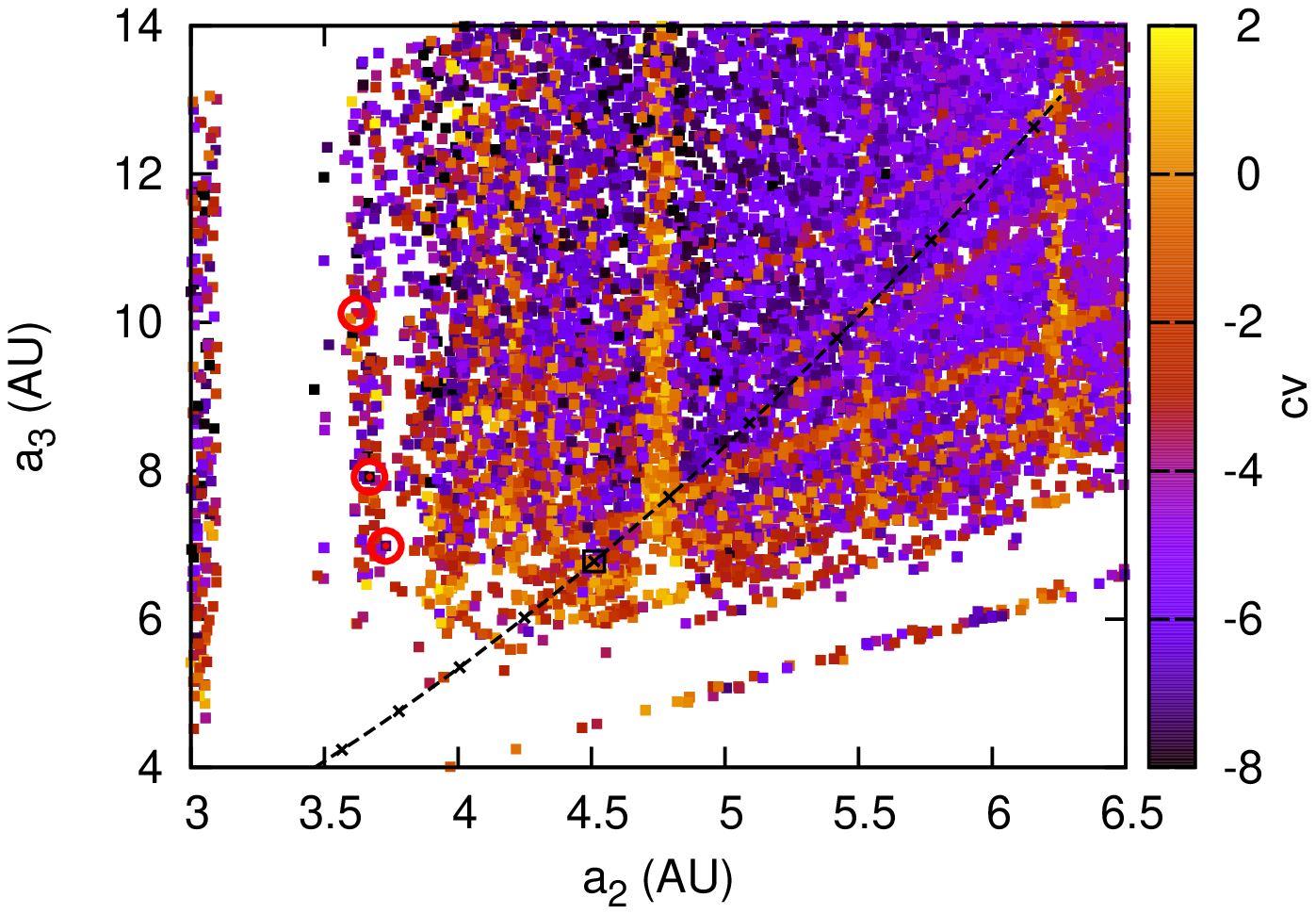}
    \includegraphics[width=0.40\textwidth,angle=-90]{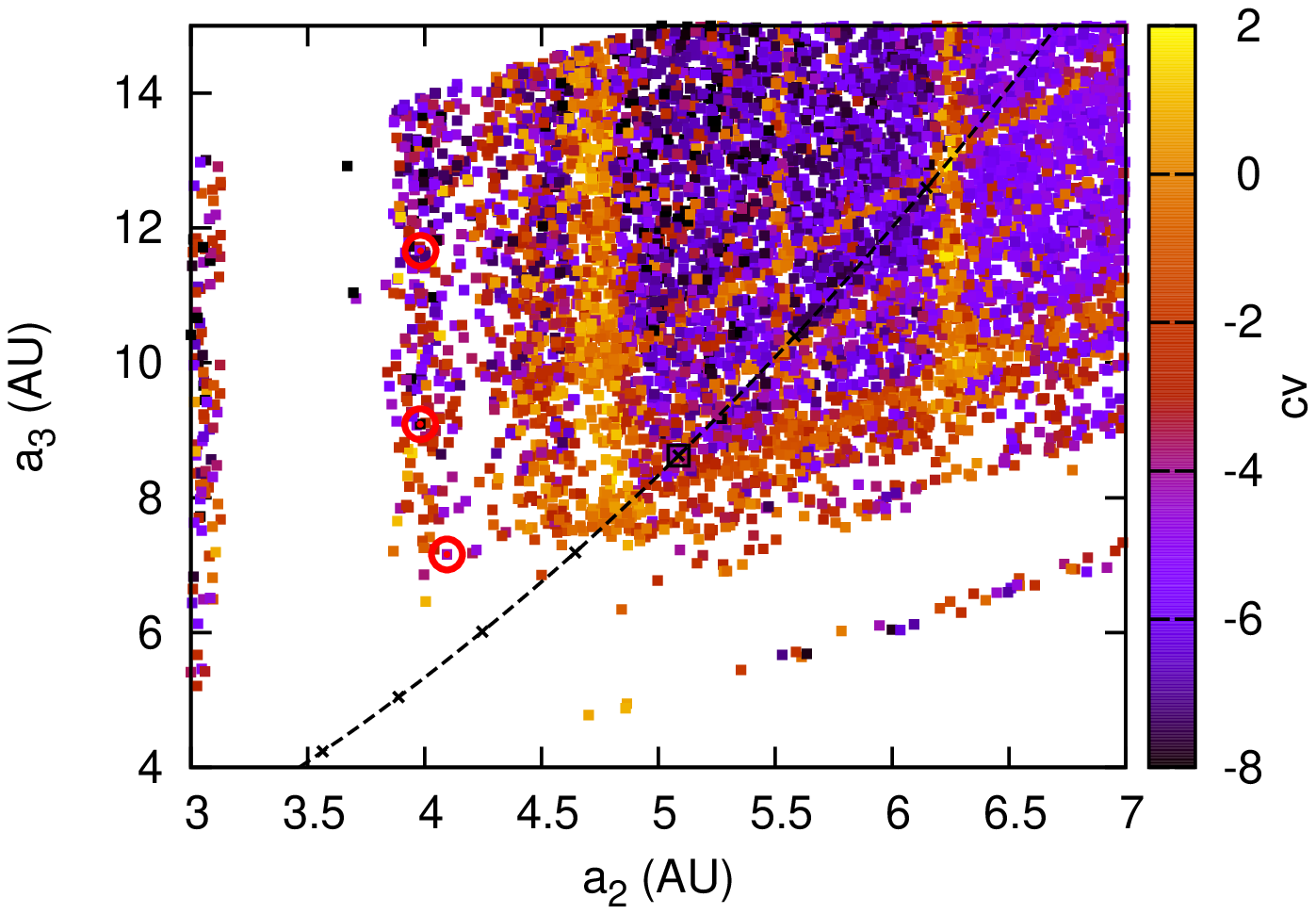}
    \caption{Stability maps for systems of 3 planets. 
    The top panel is for Neptune--size planets, the middle for 
    Saturn--size planets and the bottom for Jupiter--size planets. 
    The color coding gives the value of the chaos indicator 
    $c_v$. Brighter colors indicate unstable orbits. The 
    black dashed line represents the curve given by 
    $a_3 = a_2 + K R_H$ for different values of $K$. The crosses 
    on the curve mark the location of integer values of 
    $K$. The red circles label cases that have been integrated 
    over 5 Myr and proven to be stable. 
    The black empty squares mark where, along the K--curve, the systems 
    enter the stable zone.
   }
    \label{f3}
  \end{figure}

In Fig.\ref{f3} $c_v$ is drawn as a function of $a_2$ and $a_3$ for
the same three values of planet masses adopted in the previous section. 
We also plot the curve given by the function:

\begin{equation}
a_3  = a_2 + K R_H
\label{eq:kappa}
\end{equation}

with ticks marking increasing integer values of $K$.
Hereinafter we will call this parametrization curve the 
$K$--curve. By inspecting these plots 
we can test the robustness of the strategy of 
following the $K$--curve to explore the 
stability of three--planet systems.
The first thing that can be noted in 
each stability map
of Fig.\ref{f3} is
the dense web of resonances in the proximity of the 
instability limit, more marked in the case of more massive
planets since the resonance width is larger.
Vertical stripes with low values of $c_v$ 
correspond to MMR between the first and second planet, while 
inclined straight lines mark MMR between the second and third 
planet.

The weakness of the choice of the $K$--curve as a guiding 
parametrization to explore the stability properties of 
these systems is clearly shown in all panels of 
Fig.\ref{f3}.  For Neptune--size planets, the value of 
$K$ for which the first stable systems are met is about 
8, marked by a square in the figure. However, there are 
two critical aspects in this finding. First of all, beyond $K = 8$ 
resonances destabilize again the system until 
$K$ is about 9 and, finally, a large region of stability is met.
Second and most important aspect, 
stable systems can be found outside the $K$--curve for smaller  
$a_2$ and values of $a_3$ which lay above the $K$--curve.
Ad example, all three systems marked by 
red circles are stable over 5 Gyr (no sign of chaos in the 
orbital elements) and they approximately corresponds to 
$a_2 = a_1 + 5 R_{H1,2}$ 
and $a_3 = a_2 + K R_{H2,3}$ with $K$ ranging 
from about 11, for the case with smaller $a_3$, and beyond. 
The circled cases are representative of a large region of the phase space 
populated by stable systems that are not met and identified by 
following the $K$--curve parametrization. 
The same effect can be seen in 
the other two panels of Fig.\ref{f3}. In all these systems the 
two inner planets are more packed respect to what it would be 
predicted by the $K$--curve parametrization. 
For Saturn--size planets
(middle panel),  
the first stable systems along the $K$--curve
appear at $ K \sim 7$. 
However, 
moving farther
along the $K$--curve, 
the 2:1 MMR between the two inner planets 
is crossed and only beyond $K \sim 8$ the systems 
become stable again. 
Even in this case the red circles indicate  
systems far from the $K$--curve which 
are stable and non--chaotic over 5 Gyr and with a more 
packed inner planet
pair. The value of $K$ for the two inner
planets is about 3.5 while for the outer planet it 
is about 10 for the system closer to the $K$--curve and 
larger for the other two marked systems. 
For Jupiter--size
planets (bottom panel), the first stable systems along the $K$--curve
are found for $K$ close to 6 (squared value in panel 3, Fig.\ref{f3}), in good 
agreement with \cite{MW02, CHATTERJEE08}. Also in 
this case there is a large portion of the parameter space,
above the $K$--curve and for 
small values of $K$, where stable systems can be found.
Ad example, the 
red circles label cases that 
do not comply with the standard parametrization 
given by  Eq.~(\ref{eq:cha3})
but are stable for 5 Gyr without any trace of 
chaotic behaviour. In conclusion, the $K$--curve used
to probe the stability of 3--planet systems is misleading
since it neglects a significant amount of 
internally more packed cases which 
are stable at least for 5 Gyr and beyond. 

It is also interesting to observe that 
Trojan planets, either the two inner ones or the two outer ones, 
can coexist with a close external perturber. 
In all three panels, the inner vertical stripe represent systems 
where the two inner planets are in a 1:1 resonance. The inclined 
straight line on the bottom shows instead systems with the 
two outer planets in a Trojan configuration. In both groups there 
are cases with very low values of $c_v$ and then stable on the 
long term. 

As in Fig.\ref{f1}, a slow degradation is observed in the 
FMA analysis and higher values of $c_v$ are found for 
larger values of both $a_2$ and $a_3$. As for the case of 
two planets, this is due to the decrease of the frequencies
and eccentricities of the more detached planetary systems. 

\section[]{Summary and Conclusions}

In this paper, the dynamics of two and three equal--mass planet 
systems has been explored in detail 
close to the stability 
limit under the guidance of the frequency map analysis.
For two planet systems, the main findings concern the 
presence of stable chaos beyond the 
separation $\Delta_h = 2 \sqrt{3} R_H$ 
which makes it difficult to talk about a 'stability limit'. 
This stable chaos is limited in phase space 
and it does not lead to 
planet--planet scattering. Another aspect coming out 
from the  FMA analysis is that $\Delta_h$ depends on the 
planet mass in a way that is not fully accounted for by the 
the definition of Hill's radius. For Neptune--size planets
the value of $\Delta_h$ appears to be underestimated, 
while it is a bit overestimated for Jupiter class planets. 
A better and more complete estimate of $\Delta_h$, derived 
from long term numerical simulations, is then needed. The
FMA can be used to outline the most promising regions 
where this exploration must be undertaken.

For three--planet systems, the parametrization 
based on the relations $a_2 = a_1 + K R_{H1,2}$
and $a_3 = a_2 + K R_{H2,3}$, widely used to study the stability of
these systems, does not appear a reliable choice. As
shown by the stability maps, there is a large portion of 
systems which are stable but violate the above mentioned parametrization
and do not lay on the $K$--curve. 
A two dimensional approach is
needed to find the transition between stable and unstable 
systems that may lead to planet--planet scattering. This 
has important consequences in modeling the initial 
stages of evolution of a planetary system that has reached 
completion. The initial 
separation, marking the transition from stable to unstable 
orbits, must be estimated via numerical simulations and 
cannot be easily determined on the basis of 
semi--empirical formulas. Care must also be used when 
analysing a newly discovered system and predictions 
based only on the above mentioned formulas should 
be tested against a more accurate dynamical analysis based on 
chaos indicators and direct numerical integration. 

The models presented in this paper are limited since they 
consider equal mass planets in initially circular and 
almost coplanar orbits. More detailed explorations have
to be performed for more dynamically complex systems
where the eccentricity or mutual inclinations are not 
negligible. In this case, the reference formulas will be 
those described in \citep{donnison06}.

\section*{Acknowledgments}
I thank an anonymous referee for his useful comments that 
helped to improve the paper. 

\bibliographystyle{aa}
\bibliography{biblio}

\bsp

\label{lastpage}

\end{document}